\pdfoutput=1
\documentclass[aps,prl,onecolumn,notitlepage,howpacs,amsmath,amsmath,amssymb,amsfonts]{revtex4-1}
\usepackage{graphicx}
\usepackage{dcolumn}
\usepackage{bm}
\usepackage{color}
\usepackage{multirow}
\usepackage{natbib}
\usepackage{hyperref}

\begin{document}
\title{Non-equilibrium fluctuations and metastability arising from non-additive interactions in dissipative multi-component Rydberg gases}

\author{Ricardo Guti{\' e}rrez, Juan P. Garrahan, and Igor Lesanovsky}
\affiliation{School of Physics and Astronomy, University of
Nottingham, Nottingham, NG7 2RD, UK}
\affiliation{Centre for the Mathematics and Theoretical Physics of Quantum Non-Equilibrium Systems, University of
Nottingham, Nottingham, NG7 2RD, UK}

\begin{abstract}
We study the out-of-equilibrium dynamics of dissipative gases of atoms excited to two or more high-lying Rydberg states. This situation bears interesting similarities to classical binary (in general $p$-ary) mixtures of particles. The effective forces between the components are determined by the inter-level and intra-level interactions of Rydberg atoms.  These systems permit to explore new parameter regimes which are physically inaccessible in a classical setting, for example one in which the mixtures exhibit non-additive interactions. In this situation the out-of-equilibrium evolution is characterized by the formation of metastable domains that reach partial equilibration long before the attainment of stationarity. In experimental settings with mesoscopic sizes, this collective behavior may in fact take the appearance of dynamic symmetry breaking. 
\end{abstract}


\maketitle

\section{Introduction}

Dissipative Rydberg gases enable the exploration of a great variety of out-of-equilibrium phenomena. Dynamical effects that have been theoretically predicted include kinetic constrains \cite{lesanovsky2013}, crystallization \cite{pohl2010,van2011,Glaetzle2012,Honing2013,lechner2015}, bistability \cite{ates2012,lee2012,Hu2013}, spatial correlations and density waves \cite{Petrosyan2013}, aggregation \cite{ates2007,lesanovsky2014}, antiferromagnetic order \cite{hoening2014}, non-equilibrium phase transitions \cite{marcuzzi2014}, classical and quantum glassiness \cite{mattioli2015}, many-body entanglement \cite{Rao2014,Lee2015} and self-similarity \cite{gutierrez2015,levi2016}. Some of these phenomena, including the build-up of correlations \cite{schwarzkopf2011,schauss2012,weber2015}, crystallization \cite{schauss2015}, kinetic constraints \cite{Valado2016}, aggregation \cite{schempp2014,Urvoy2015} and bimodality \cite{carr2013,schempp2014,Malossi2014} have already been observed, which highlights the power of Rydberg gases for investigating non-equilibrium quantum dynamics.

While {\it single-component} systems, where one Rydberg transition is driven, have been the focus of many efforts, the dynamics of {\it multi-component} Rydberg gases ---i.e. systems with atoms excited to several Rydberg states--- remains largely unexplored. As recent experiments are starting to probe multiple Rydberg states \cite{gunter2013, Bettelli2013,Gorniaczyk2014,Barredo2015}, it is important to achieve some understanding of the collective phenomena that can be expected to be found in such systems. One can anticipate that several competing length scales will arise from the interplay between intra-level and inter-level interactions (i.e. the interactions between atoms excited to the same or different levels, respectively).  Indeed, a few theoretical studies have started exploring this competition  \cite{levi2015,qian2015}.

The study of multi-component systems may help to further the strong analogies between the dynamics of dissipative Rydberg gases and soft-matter systems \cite{lesanovsky2013,lesanovsky2014,mattioli2015}. This connection ultimately originates from the Rydberg blockade effect \cite{jaksch2000,lukin2001}, whereby an excitation of a given atom prevents that of neighboring atoms, an effect reminiscent of the excluded-volume interactions characteristic of soft-matter systems such as liquids and colloids \cite{binder2011}.  These systems are often mixtures composed of more than one kind of particle, as such dispersity can give rise to interesting collective effects that are not present in the monodisperse case, see e.g. \cite{Abraham2008}.  Furthermore, it is common when modelling soft matter computationally to consider ``non-additive'' mixtures, meaning mixtures where the cross interactions between different kinds of particles are not given by those between similar kinds: for example, if particles $A$ and $B$ interact among themselves with typical distances $\sigma_A$ and $\sigma_B$, respectively, the distance for cross interaction is such that $\sigma_{AB} \neq (\sigma_A + \sigma_B)/2$, as in e.g. Ref. \cite{kob1995}. An illustration is given in Fig. \ref{fig1} (a).    While non-additive interactions are unphysical in a classical setting where particles interact by excluded volume or similar effects, they are used to increase frustration in model liquids, thus precluding crystallization and promoting glass formation. In dissipative Rydberg gases the non-additivity of inter-atomic interactions is an experimentally realizable physical feature. Despite its quantum origin, such non-additivity survives in an effectively classical limit,  giving a new handle on experimentally realizable binary (or generally $p$-ary) mixtures.

In this work, we elucidate the physics of multi-component  dissipative Rydberg gases far from equilibrium. We first develop a general theory for the dynamics of systems with any number of components extending an approach that has been extensively validated in the one-component case \cite{Valado2016,Urvoy2015}. We then perform an idealized numerical study, where different interaction strengths lead to a variety of  length scales giving rise to strikingly different dynamical regimes. The phenomenology that emerges from non-additive interactions is characterized by the formation of domains, which are homogeneously populated by excitations of a given component when inter-level interactions dominate, and show an alternation of components in the opposite case. Homogeneous domains reach partial equilibration when detailed balance is achieved for the dominant atomic transition, leading to metastable behavior. In experimental settings with mesoscopic sizes, these domains will appear as non-equilibrium symmetry-broken states.

\section{Theory: Effective dynamics in the limit of strong dissipation}

\begin{figure}[t]
\includegraphics[scale=0.18]{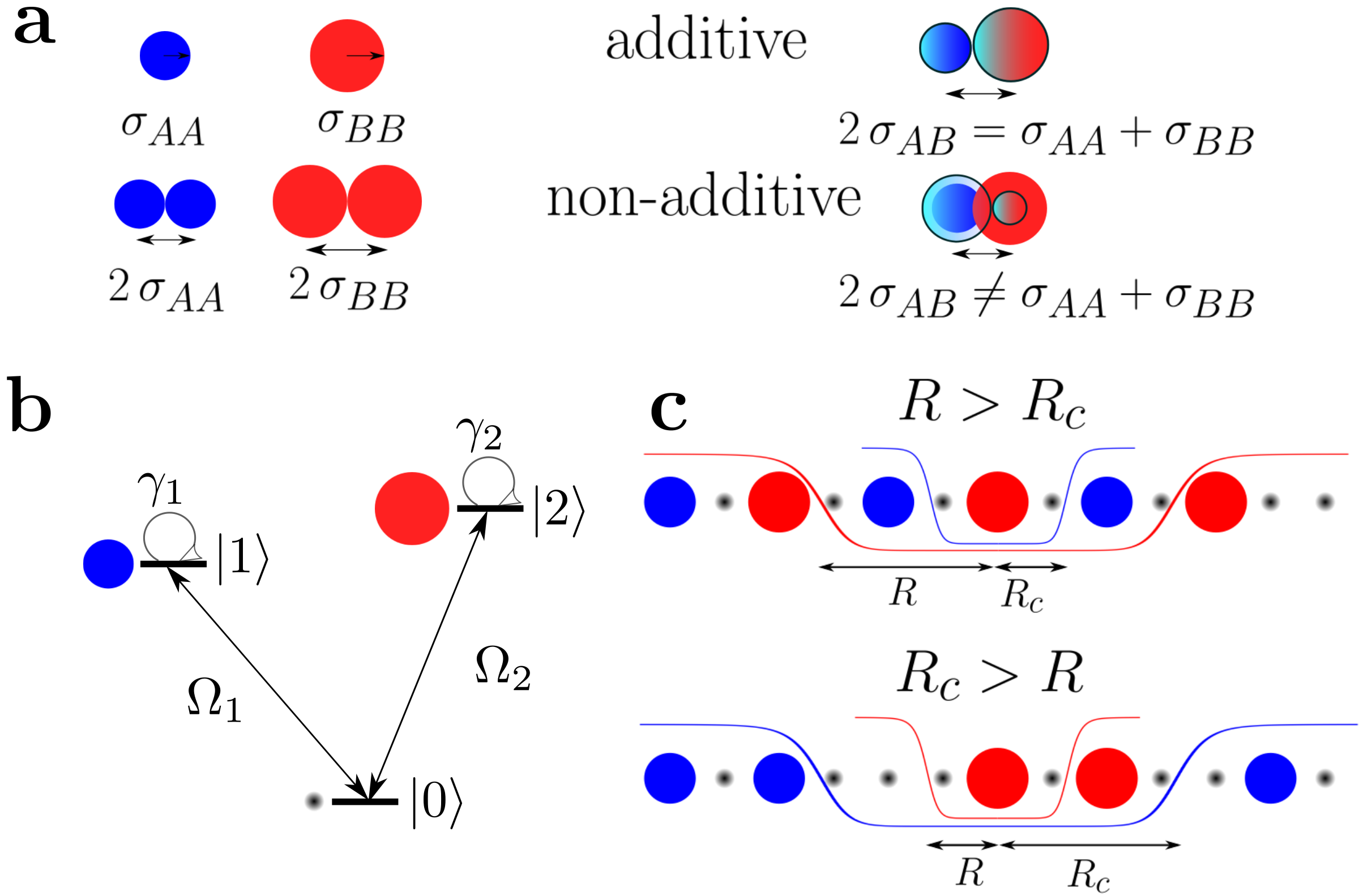}
\vspace{-0.3cm}
\caption{ {\sf \bf Non-additive interactions, energy level scheme, and competing length scales.}  (a) Additive and non-additive interactions (see text for definition). (b) Atomic energy levels, dephasing rates, and laser-driven transitions. (c) Competing length scales $R$ and $R_c$ for intra- and inter-level interactions, respectively: $R>R_c$ ($R<R_c$) leads to alternating patterns (homogeneous regions).}\label{fig1}
\end{figure}

We consider a system of $N$ atoms, each of which can be in one of $p+1$ levels, the ground state $|0\rangle$,  and $p>1$ Rydberg states $|1\rangle$, $|2\rangle$, \ldots $|p\rangle$, with energies $E_0 < E_1 <E_2 < \cdots < E_p$. See  Fig. \ref{fig1} (b) for an ilustration of the two-component case. Atoms in the Rydberg states  $|s\rangle$ and  $|s^\prime\rangle$ at positions ${\bf r}_k$ and ${\bf r}_m$ interact through a power-law potential $V_{km}^{ss^\prime} = C_\alpha^{ss^\prime}/|{\bf r}_k - {\bf r}_m|^\alpha$ with exponent $\alpha$. For simplicity, we denote the intra-level interactions by $V_{km}^{s}$ instead of  $V_{km}^{ss}$. The value of the coefficients $C_\alpha^{ss^\prime}$ depends on the specific structure of the atomic spectrum and can be controlled through e.g. electric field induced F\"{o}rster resonances \cite{gorniaczyk2015} or microwave dressing \cite{Marcuzzi2015,sevinccli2014}. Typically encountered exponents are $\alpha=6$ (van der Waals interaction) and $\alpha=3$ (dipole-dipole interaction) \cite{saffman2010}. Each of the Rydberg states is resonantly coupled to the ground state by a laser field, and affected by dephasing noise \cite{schempp2014,Urvoy2015,Valado2016}. The dynamics of the system is governed by a Master equation of Lindblad form $\partial_t \rho = \mathcal{L} \rho + \mathcal{D}(\rho)$ \cite{robert2012}. The coherent part $\mathcal{L} \rho = -i[H_0+H_1,\rho]$ includes an interaction Hamiltonian
\begin{equation}
H_0 =  \sum_{s=1}^p \sum_{k < m} \left[ V_{km}^{s} n_s^{(k)} n_s^{(m)} + \sum_{s^\prime \neq s} V^{s s^\prime}_{km} n_s^{(k)} n_{s^\prime}^{(m)}\right],
\label{h0pspec}
\end{equation}
and a driving term $H_1 = \sum_{s=1}^p \Omega_s \sum_{k=1}^N \sigma_{sx}^{(k)}$. Here, $n_s^{(k)}=\left|s\right>_k\!\left<s\right|$, $\sigma_{sx}^{(k)}=\left|s\right>_k\!\left<0\right| + \left|0\right>_k\!\left<s\right|$ and $\Omega_s$ is the Rabi frequency of the transition between $\left|0\right>$ and $\left|s\right>$.  The dissipator is given by  $\mathcal{D}(\rho) = \sum_{s=1}^p  \gamma_s \sum_{k=1}^N\!\left(n_s^{(k)} \rho\, n_s^{(k)} - \frac{1}{2} \left\{n_s^{(k)},\rho\right\}\right)$, where $\gamma_s$ is the dephasing rate of $\left|s\right>$ w.r.t. $\left|0\right>$. 
Atomic decay is not considered, as we are especially interested in the short time dynamics that has been probed in experiments \cite{Urvoy2015, Valado2016}. We deliberately focus on a situation where exchange interactions can be omitted, which can be achieved by a specific choice of Rydberg states \cite{gorniaczyk2015}.

In the limit of strong dissipation, $\Omega_s \ll \gamma_s$, which is relevant in a number of experimental settings \cite{Urvoy2015,Valado2016}, the time evolution is governed by an stochastic dynamics along the classical states represented in $\mu = \textrm{diag}(\rho)$ \cite{lesanovsky2013,cai2013}. While the effective equations of motion of the multi-component Rydberg gas are crucially important for the rest of the paper, and simple enough as to provide physical insight into the phenomenology that is numerically observed (which would be very hard to infer from the quantum master equation), their derivation is relatively lengthy. We therefore include here only the main results, and give the technical details in Appendix A for the interested reader. The resulting rate equations are
\begin{equation}
\partial_t \mu = \sum_{s=1}^p \frac{4\Omega_s^2}{\gamma_s} \sum_k \Gamma_{s}^{(k)} \left[\sigma_{sx}^{(k)} \mu \sigma_{sx}^{(k)} -  \mathcal{I}_s^{(k)} \mu \right],
\label{effectiveintpspec}
\end{equation}
where $ \mathcal{I}_s^{(k)} = n_{s}^{(k)} + |0\rangle_k \langle 0|$ projects on the subspace spanned by the ground state and the excited state $|s\rangle$ of site $k$. For simplicity, we  assume that the atoms sit in the sites of a chain with lattice constant $a$. A transition involving the excited level $|s\rangle$ at site $k$, whether it is an excitation or a de-excitation, occurs with a rate 
\begin{equation}
\frac{1}{\Gamma_{s}^{(k)}} = 1\! +\! \left[\sum_m \frac{\left(R_{s}\right)^\alpha\! n_{s}^{(m)}\!+\! \sum_{s^\prime \neq s}\left(R^{ss^{\prime}}_{s}\right)^\alpha\! n_{s^\prime}^{(m)}}{|\hat{\bf r}_k - \hat{\bf r}_m|^\alpha}\right]^2\!,
\label{ratesintpspec}
\end{equation}
where  $\hat{\bf r}_k = {\bf r}_k/a$. The relevant length scales are given by the {\it intra-level},  $R_{s}=a^{-1}\left[2C^{s}_\alpha/\gamma_{s}\right]^{1/\alpha}$, and the {\it inter-level} interaction parameters, $R_{s}^{s s^\prime}=a^{-1}[2C^{s s^{\prime}}_\alpha/\gamma_{s}]^{1/\alpha}$, which are the reduced distances at which the appearance of excitations of a given component correlate different sites. As in classical mixtures of particles (liquids, colloids, etc.) several components coexist and their interactions are characterized by different typical length scales depending on the components involved.

Experiments typically probe the dynamics starting from an initial state where all atoms are in the ground state, and this will be our choice as well. At the initial stages distant excitations to any level occur independently of each other with a rate that is $\mathcal{O}(1)$. This gives an ``initial seed'' for the correlated dynamics: as soon as the distance between excitations becomes comparable with $R_{s}$ and/or $R_{s}^{s s^\prime}$,  the second term in Eq. (\ref{ratesintpspec}) strongly correlates the atoms, and the transitions between the ground state and a particular level become less likely due to the presence of nearby excited particles [see Fig. \ref{fig1} (c)].

\section {Phenomenology: Numerical results} 

We next turn to a numerical exploration of the phenomenology that emerges in multi-component Rydberg gases. As the dynamics is rich in collective effects, we start from the simplest possible case of a two-component system with symmetric interaction parameters, $R_{1} = R_{2} \equiv R$ and $R_{1}^{12} = R_{2}^{12} \equiv R_c$. The expression in brackets on the rhs of Eq. (\ref{ratesintpspec}) then becomes $\sum_m \left( R^\alpha\,n_{1}^{(m)}+R_{c}^\alpha n_{2}^{(m)}\right)/|\hat{\bf r}_k - \hat{\bf r}_m|^\alpha$ for transitions between $|0\rangle_k$ and $|1\rangle_k$, and an equivalent expression for the transition between $|0\rangle_k$ and $|2\rangle_k$ is obtained by swapping $n_{1}^{(m)}$ and $n_{2}^{(m)}$. In keeping with the aim to simplify the parameter space as much as possible, we further assume $\Omega_1^2 / \gamma_1 = \Omega_2^2 / \gamma_2$, and rescale the time variable by $\Omega_1^2 / \gamma_1$. We focus our study on three generic cases: ({\it i}$\,$) $R>R_c$, ({\it ii}$\,$) $R\sim R_c$ and ({\it iii}$\,$) $R<R_c$. Cases  ({\it i}$\,$) and  ({\it iii}$\,$) are examples of non-additive interactions, $R \neq R_c$. Such interactions are of interest in the theoretical study of complex and glassy dynamics in classical mixtures, but their experimental realization remains challenging in those contexts, whereas they appear generically in Rydberg gases.  We expect that in case  ({\it i}$\,$) the excitations of one component will be surrounded by excitations of the other component, in an anticorrelated pattern, as in the upper panel of Fig. \ref{fig1} (c). By analogy, in case ({\it iii}$\,$), one expects the clustering of excitations of a given component, as in the lower panel of Fig. \ref{fig1} (c).  

\begin{figure}[h]
\hspace{-0.27cm}
\includegraphics[scale=0.18]{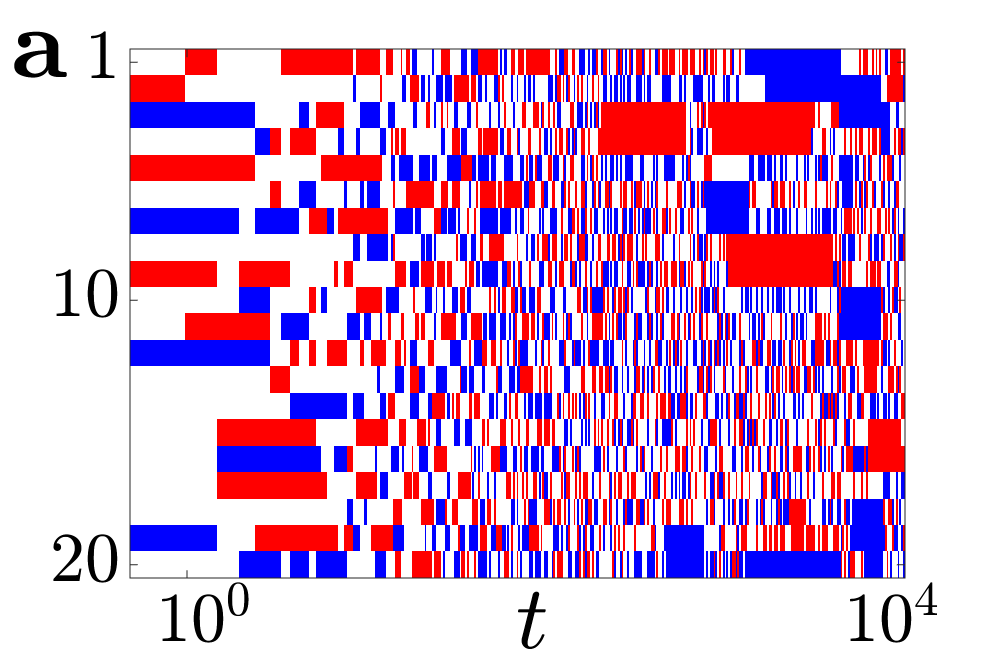}\hspace{-0.1cm}
\includegraphics[scale=0.18]{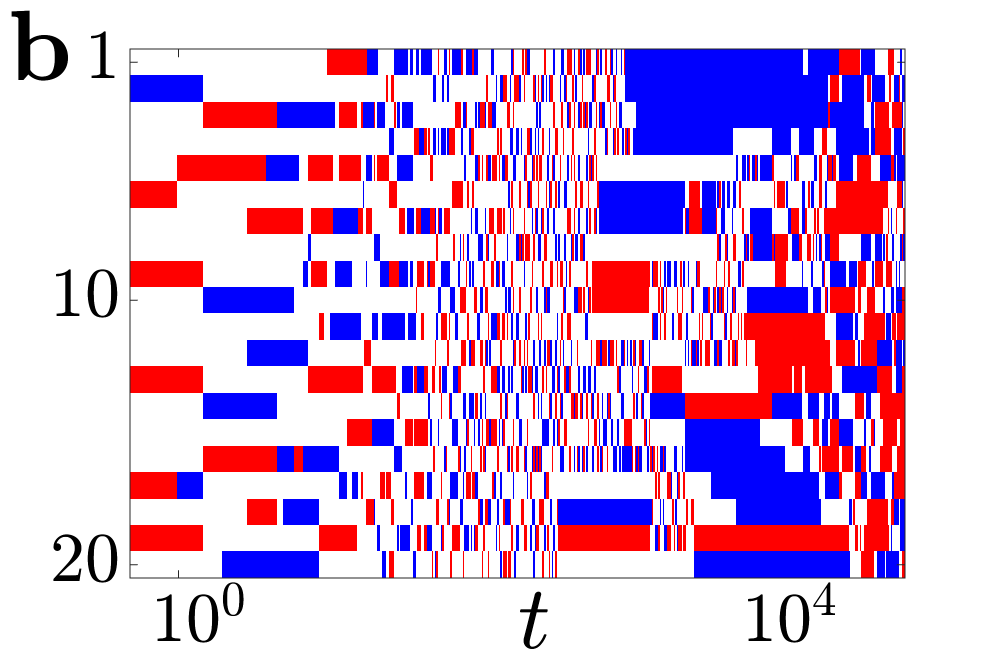}\\
\hspace{-0.27cm}
\includegraphics[scale=0.18]{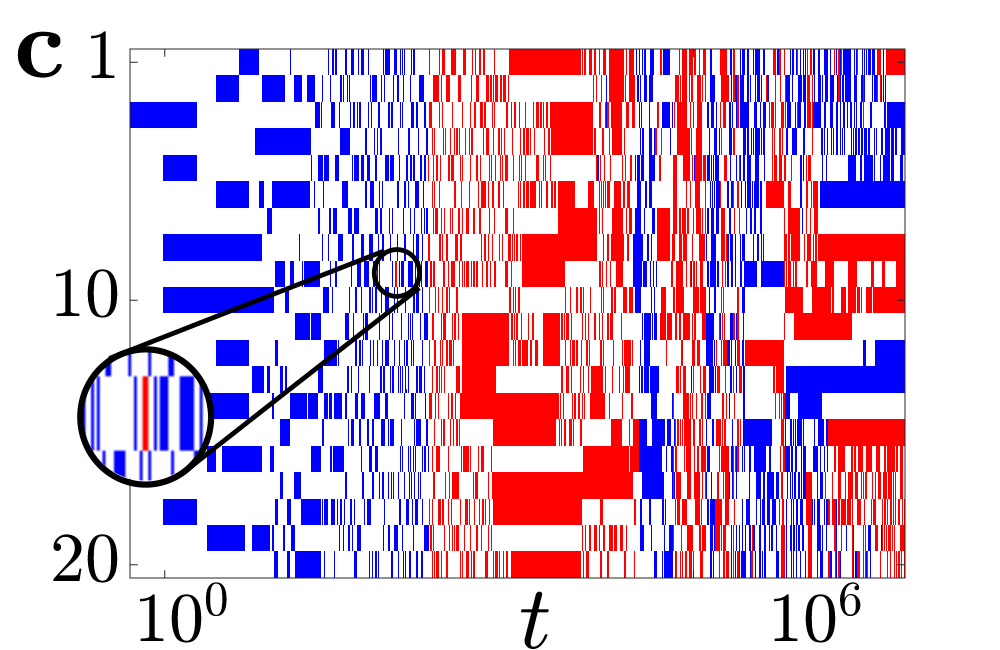}\hspace{-0.1cm}
\includegraphics[scale=0.18]{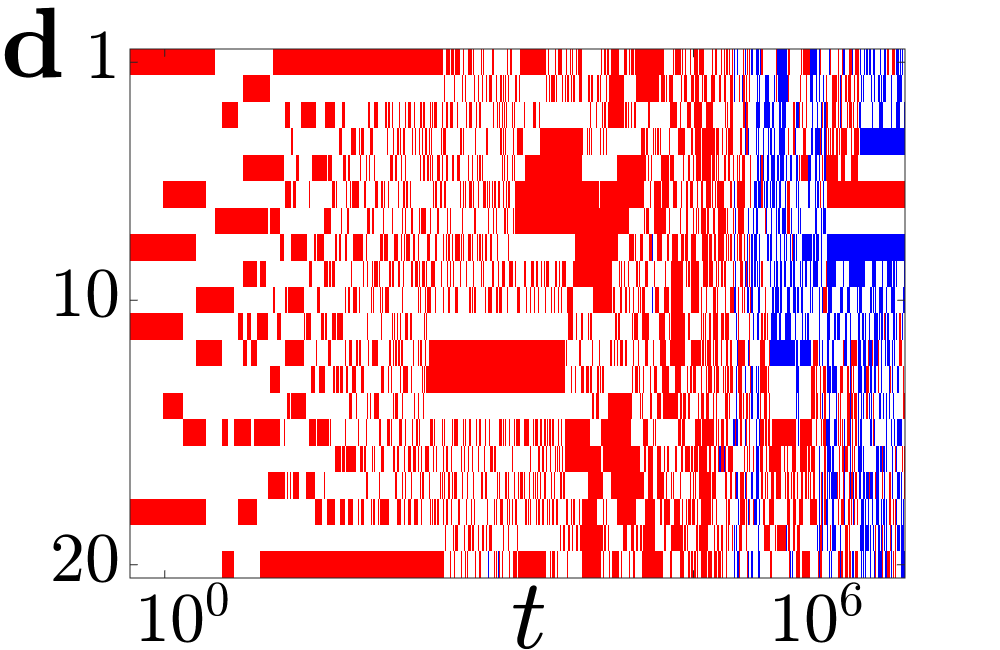}\\
\vspace{-0.2cm}
\caption{{\sf \bf Representative trajectories for different inter-level interaction parameter values.}
Representative trajectories for $R=2$ and $R_c = 1/2$ (a),  $R_c = 2$ (b), and $R_c = 8$ (c, d). The appearance of an ``intruding'' excitation has been magnified in (c). Blue corresponds to $|1\rangle$, red to $|2\rangle$, and white to ground state atoms.}\label{fig2}
\end{figure}

This phenomenology is indeed observed using kinetic Monte-Carlo simulations in a 1D chain of van der Waals-interacting atoms ($\alpha = 6$). We focus on a mesoscopic system of size $N=20$, as such sizes are accessible by current experiments, and use periodic boundary conditions to prevent uncontrolled boundary effects.  In Fig. \ref{fig2} we show representative trajectories for cases ({\it i}$\,$), ({\it ii}$\,$) and ({\it iii}$\,$) for fixed $R=2$ and varying $R_c$, where an atom appears in blue if it is in the excited state $|1\rangle$, in red if it has been excited to $|2\rangle$, and in white if it is in the ground state. Analogous results for four components are presented in Appendix B. For $R_c = 1/2$ ($R > R_c$), the excitation pattern forms something that can be described as {\it heterogeneous domains} of alternating excitations of one and the other component, with some defects [Fig. \ref{fig2} (a)]. For $R_c=8$ ($R < R_c$), where the proximity of heterogeneous neighbors is penalized, large {\it homogeneous domains} (i.e. regions where there are only excitations of a given component) are seen to exist throughout most of the non-equilibrium evolution of the system [Fig. \ref{fig2} (c) and (d)]. Whether one sees a homogeneous domain of one or the other component depends on the small imbalances that may occur at the initial stages of the process. Indeed, in Fig. \ref{fig2} (c) at some point, the appearance of ``intruding'' excitations (one of them is magnified) leads to the replacement of a large component $|1\rangle$ domain by a similar one of component $|2\rangle$. In other trajectories, like that shown in  Fig. \ref{fig2} (d), domains of a given component dominate throughout the non-equilibrium regime.  As for the situation in which $R_c = 2$ [$R = R_c$, Fig. \ref{fig2} (b)], corresponding to additive interactions, excited atoms are as likely to be found close to excitations of either component throughout the evolution of the system. Indeed, the components act as labels that permit to distinguish different types of excitations, but they have no dynamical consequences. This is in stark contrast to situations in which the interactions are non-additive, where (as shown above) the configurations that emerge are highly dependent on the components of the excitations. As in mixtures of classical particles, non-additivity brings richness into the dynamics.

\begin{figure}[h]
\includegraphics[scale=0.16]{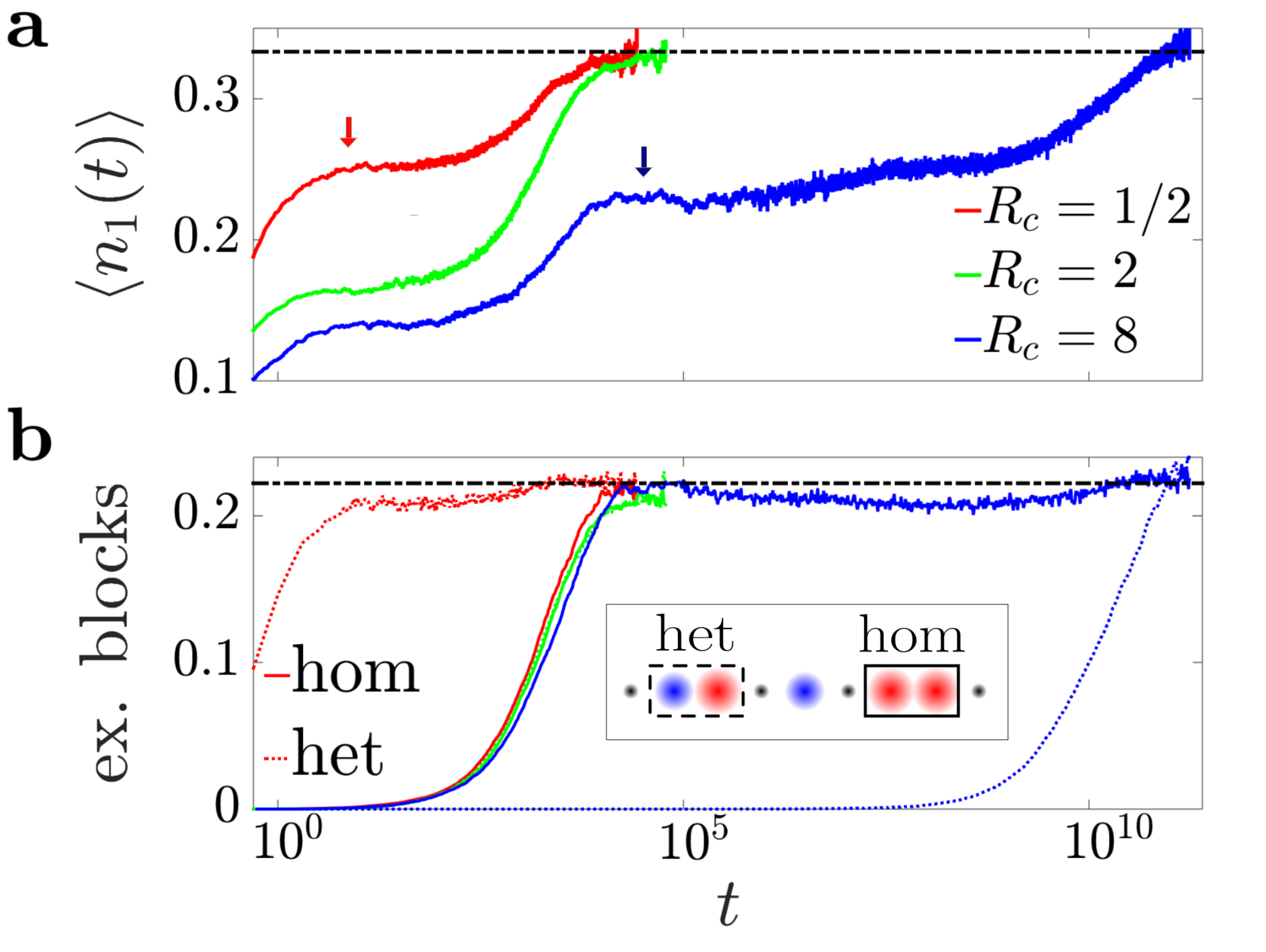}\vspace{-0.3cm}
\caption{{\sf \bf Density of excitations and excitation blocks.}
(a) Density of component $|1\rangle$ atoms  for $R=2$ and  $R_c = 1/2, 2, 8$. Arrows indicate the times around which the density reaches a plateau (as explained in the text) for $R_c=1/2$ (red) and $R_c=8$ (blue). (b) Density of homogeneous (continuous lines) and heterogeneous (dotted lines) excitation blocks [same color coding as in (a)]. Black horizontal lines indicate the stationary values. Averages based on $2000$ trajectories.}\label{fig3}
\end{figure}

With increasing time, the lattice fills with more and more excitations. Eventually these highly structured configurations disappear and the system settles into the stationary state of the master equation, which is proportional to the identity, $\rho_\mathrm{st} \equiv (p+1)^{-N} \otimes_k \mathbb{I}_k$. Accordingly, in the effective dynamics the average number of atoms in each level becomes $N/(p+1)$, as can be seen from Eq. (\ref{effectiveintpspec}).  In Fig. \ref{fig3} (a) we show $\langle n_1(t) \rangle \equiv (1/N) \sum_k \langle n_1^{(k)}(t) \rangle$, i.e. the density of atoms in the excited state $|1\rangle$, as a function of time. This observable gives us some important information of the generic aspects of the classes of dynamics illustrated in Fig. \ref{fig2} for specific realizations. We again fix $R = 2$, and look at $R_c = 1/2, 2$ and $8$. The excitation density for the situations corresponding to the two extreme values of $R_c$ increases until it reaches a long plateau which has been highlighted with vertical arrows in the figure. The origin of these plateaus will be clarified below. Much later another increase leads the system towards the sationary state (see the black horizontal line). While the results  for $\langle n_2^{(k)}(t)\rangle$ are identical, single trajectories fluctuate strongly, a situation reminiscent of dynamic symmetry breaking (see, e.g., \cite{evans1995,kim2006}).

To gain insight into the relaxation behavior reported in  Fig. \ref{fig3} (a), and especially on the type of configurations that occur at different stages of the dynamics, it is useful to complement the study of the time evolution of the density of excitations with that of an observable that can help distinguish between different local patterns of excitations. For this purpose, we focus on the density of {\it excitation blocks}, i.e. excited atoms whose right neighbors are also excited. We consider separately homogeneous and heterogeneous blocks, which are made up of same-component or different-component excitations respectively. The former type is shown enclosed in a continuous-line box and the latter in a dashed-line box in the inset of Fig. \ref{fig3} (b), where we show the density of homogeneous (continuous line) and heterogeneous blocks (dotted line). The block density in the stationary state is indicated by a black horizontal line. This is $\langle n_1^{(i)} n_1^{(i+1)}\rangle/N + \langle n_2^{(i)} n_2^{(i+1)}\rangle/N = 2/(p+1)^2$ for homogeneous blocks, and the same value can be easily seen to apply to heterogenous blocks. For $R_c = 1/2$ the density reaches the plateau in Fig. \ref{fig3} (a) at the time the concentration of heterogeneous blocks gets close to the equilibrium value, and the final push into stationarity corresponds to an equivalent move on the part of the concentration of homogeneous blocks. This corresponds to a rapidly achieved alternating pattern of excitations, which persists for long times until it finally relaxes into the stationary state.  For $R_c = 8$, we see the opposite behavior: the plateau is reached first when the homogeneous blocks attain the equilibrium value, and stationarity is achieved after a long wait when the heterogeneous blocks reach that value as well. The interpretration is analogous to that of the $R_c = 1/2$ case, but now the domains that appear at the time the plateau is reached are homogeneous. In which case it takes shorter or longer for the homogeneous or the heterogeneous blocks to reach the equilibrium value can of course be inferred from the rates in Eq. (\ref{ratesintpspec}). The case where $R_c = R =  2$ unsurprisingly shows a simultaneous equilibration of both types of blocks, and therefore stationarity is reached without an intermediate plateau.

These results suggest that the domain structure remains in place for very long times before reaching stationarity. To clarify this we consider the order parameter
\begin{equation} 
P_{+}(t) \equiv \frac{1}{N}\sum_{k=1}^N \left[n_1^{(k)}(t) - n_2^{(k)}(t)\right],
\label{pplus}
\end{equation}
for the study of homogeneous domains. Bimodal distributions of this parameter indicate very strong dominance of one of the two components, while a narrow unimodal distribution that peaks at zero indicates the existence of configurations where both components are strongly mixed. We further define a similar parameter that assigns an alternating sign to consecutive excitations along the chain for the study of heterogeneous domains $P_{-}(t) \equiv \frac{1}{N} \sum_{k\in \mathcal{E}} (-1)^{N_k} \left[n_1^{(k)}(t) - n_2^{(k)}(t)\right]$, where $\mathcal{E}$ is the positionally-ordered set of the excitations in the chain, and $N_k$ is the position of site $k$ in  $\mathcal{E}$ (i.e., if $\mathcal{E} = \{1, 4, 9, 16,\ldots\}$, $N_1 = 1$, $N_4 = 2$, and so on). 

\begin{figure}[t]
\includegraphics[scale=0.17]{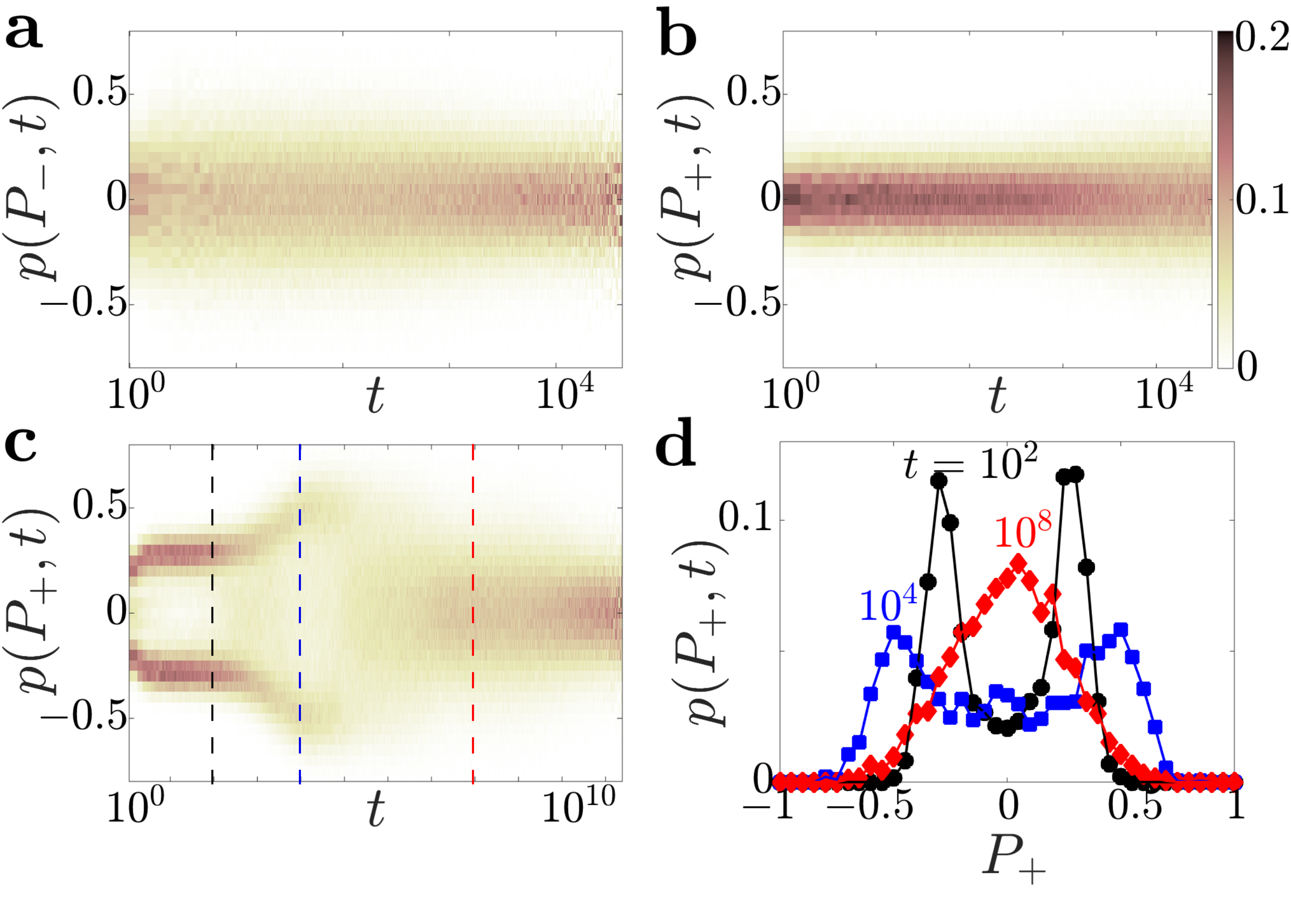}
\vspace{-0.5cm}
\caption{{\sf \bf Probability distribution of order parameters $P_{\pm}$ for different values of $R_c$.}
(a) Distribution of $P_-(t)$ for $R=2$ and $R_c=1/2$. (b,c) Distribution of $P_+(t)$ for $R= 2$ and $R_c=2$ (b) or $R_c=8$ (c). (d) Distributions shown in (c) at $t=10^2$ (black), $10^4$ (blue) and $10^8$ (red) (see vertical lines). Histograms contain $2000$ trajectories.}\label{fig4}
\end{figure}

The probability distribution of $P_{\pm}$ across time for $R = 2$, $R_c = 1/2, 2$ and $8$ is shown in Fig. \ref{fig4} (a), (b) and (c), respectively. For $R_c=1/2$ [Fig. \ref{fig4} (a)], $P_-(t)$ has a relatively wide distribution that narrows down as the system approaches stationarity, indicating the loss of order. The presence of defects makes the distribution unimodal even for short times. For $R_c=2$ [Fig. \ref{fig4} (b)], $P_+(t)$ is narrowly distributed around zero, as the occurrence of both components is equally likely in all realizations. This case can be analyzed with $P_-(t)$ as well, yielding very similar results (not shown). A richer phenomenology occurs when $R_c > R$  [$R_c = 8$, Fig. \ref{fig4} (c)], with a clearly bimodal distribution throughout the non-equilibrium evolution of the system. Initially, two branches are formed symmetrically around zero, separated by a region of very low probability of occurrence. This corroborates the role of the initial seed in leading the system to domains of either component. The two peaks of $P_+(t)$ separate more and more until they saturate. Later on, the domain structure starts crumbling upon the appearance of excitations of the non-dominant component. This is illustrated in Fig. \ref{fig4} (d), where the curves corresponding the distributions shown in (c) at times $t=10^2, 10^4$ and $10^8$ are shown.

The saturation value for $R_c > R$ is $|P_+|\simeq 0.5$, which corresponds to a {\it metastable state}, as we now explain. In the two-component case, the right-hand side of Eq. (\ref{effectiveintpspec}) contains two terms for each site. Within a homogeneous domain of, say, component $|1\rangle$, $\Gamma_2^{(k)}\ll\Gamma_1^{(k)}$, so for times shorter than $1/\Gamma_2^{(k)}$  transitions between $|0\rangle$ and $|1\rangle$ dominate. The corresponding term reaches a ``partial equilibrium'' when $\sigma_{1x}^{(k)} \mu \sigma_{1x}^{(k)} -  \mathcal{I}_1^{(k)} \mu$ is as likely to create excitations as de-excitations, i.e. when there are as many atoms in the ground state as in state $|1\rangle$. Indeed, this state, in which detailed balance is satisfied for one of the transitions between the ground state and an excited state, would correspond to the stationary dynamics of a single-component Rydberg gas. In multi-component systems, however, excitations of the non-dominant component have to appear eventually in order for the system to reach the true stationary state, as shown in Fig. \ref{fig4} (c) and (d). A similar behavior is observed in four-component systems (see Appendix B). The reader should note that such metastable states may not be achieved in experiments starting from an empty initial state if the times required to reach them exceed the lifetimes of the atoms. Starting from densely populated initial states can be helpful in probing this metastability.

\section{Conclusions} 

We have derived an effective theory for a multi-component Rydberg gas in the presence of noise. For non-additive interactions, the emerging dynamics displays a domain structure that depends sensitively on the initial excitations.  For large inter-species interactions this leads to a metastable dynamics when partial equilibration is reached for the dominant component, which corresponds to the stationary state of a single-component system where only that Rydberg transition is driven. To our knowledge, this could be the first system that is experimentally accessible in which non-additive interactions of the kind that are considered in classical mixtures of particles for the study of metastable dynamics can be naturally implemented, and are indeed expected to occur generically. Whether the phenomenology persists at the qualitative level when the dissipation is only moderately strong or even weak compared to the driving, as has been recently shown to occur in the case of single-component Rydberg gases \cite{levi2016}, is an interesting question that remains to be studied, as is the general role of quantum fluctuations \cite{marcuzzi2016}. The possibility that the (to some extent) tunable exchange interaction of Rydberg gases \cite{gorniaczyk2015} can open up new relaxation pathways in multi-component systems will be explored in the future.

\section{Appendix A. Derivation of the effective equations of motion}

We consider a gas of $N$ atoms in a lattice. The ground state $\left|0\right>$ of each atom is resonantly coupled by laser fields to the Rydberg states $\left|1\right>, \left|2\right>, \ldots, \left|p\right>$ (with energies such that $E_0 < E_1 < \cdots < E_p$). The Master equation is then given by $\partial_t \rho = \mathcal{L}_0 \rho + \mathcal{L}_1 \rho$, where $\mathcal{L}_0$ contains the interaction Hamiltonian and the dissipator, and $\mathcal{L}_1$ gives the time evolution due to the driving. For the derivation below, where $\mathcal{L}_1$ will be treated as a perturbation, this is more convenient than the more physical decomposition into a coherent part and a dissipator that is used in the main text. The Liouvillian superoperator $\mathcal{L}_0$ is defined as
\begin{equation}
\mathcal{L}_0 \rho = -i[H_0,\rho] + \sum_{s=1}^p  \gamma_s \sum_{k=1}^N\left(n_s^{(k)} \rho\, n_s^{(k)} - \frac{1}{2} \left\{n_s^{(k)},\rho\right\}\right),
\label{Al0pspec}
\end{equation}
where $n_s^{(k)}=\left|s\right>_k\!\left<s\right|$ and $\gamma_s$ is the dephasing rate of $\left|s\right>$ w.r.t. $\left|0\right>$. Atoms in the Rydberg states  $|s\rangle$ and  $|s^\prime\rangle$ at positions ${\bf r}_k$ and ${\bf r}_m$ interact through a power-law potential $V_{km}^{ss^\prime} = C_\alpha^{ss^\prime}/|{\bf r}_k - {\bf r}_m|^\alpha$ with exponent $\alpha$. For simplicity we denote the intra-level interactions by $V_{km}^{s}$ instead of  $V_{km}^{ss}$. As a result, the Hamiltonian $H_0$ can be written as
\begin{equation}
H_0 =  \sum_{s=1}^p \sum_{k < m} \left[ V_{km}^{s} n_s^{(k)} n_s^{(m)} + \sum_{s^\prime \neq s} V^{s s^\prime}_{km} n_s^{(k)} n_{s^\prime}^{(m)}\right].
\label{Ah0pspec}
\end{equation}
The superoperator $\mathcal{L}_0$ therefore consists of a Hamiltonian part and a dissipator whose individual terms commute. Additionally, we have the driving term, which in the rotating-wave approximation becomes
\begin{equation}
\mathcal{L}_1 \rho = -i \sum_{s=1}^p \Omega_s \sum_{k=1}^N [\sigma_{sx}^{(k)},\rho]
\label{Al1pspec}
\end{equation}
where $\sigma_{sx}^{(k)}=\left|s\right>_k\!\left<0\right| + \left|0\right>_k\!\left<s\right|$ and $\Omega_s$ is the Rabi frequency of the coupling between $\left|s\right>$ and $\left|0\right>$.

Our aim is to derive the effective dynamics in the limit of strong dissipation, $\Omega_s \ll \gamma_s$. We start by working out the effect of the dissipator on the dynamics.  Using the notation, $\mathcal{L}_{0,d}^k = \sum_{s=1}^p \left[ \gamma_s \left(n_s^{(k)} \rho\, n_s^{(k)} - \frac{1}{2} \left\{n_s^{(k)},\rho\right\}\right)\right]$, we write 
\begin{equation}
e^{\mathcal{L}_0 t} \rho = e^{-i H_0 t} \left[ \bigotimes_m e^{\mathcal{L}_{0,d}^m t} \rho \right] e^{i H_0 t}.
\label{Aactiondissip}
\end{equation}
The first-order contribution of the action of $e^{\mathcal{L}_{0,d}^k t} = \mathbb{I} + \mathcal{L}_{0,d}^k \, t + 1/2!\, (\mathcal{L}_{0,d}^k)^2 \, t^2 + \mathcal{O}(t^3)$ on the density operator is
\begin{eqnarray}
&\mathcal{L}_{0,d}^k\, \rho_{ij}^{(k)} \, t\, &= \left( \begin{array}{ccccc}
0 & -\frac{1}{2}(\gamma_{p-1}+\gamma_p)\, t\, \rho_{p(p-1)}^{(k)} & \cdots &  -\frac{1}{2} \gamma_p\, t\, \rho_{p0}^{(k)}  \\
-\frac{1}{2}(\gamma_{p-1}+\gamma_p)\, t\, \rho_{(p-1)p}^{(k)} & 0 & \cdots &  -\frac{1}{2} \gamma_{p-1}\, t\, \rho_{(p-1)0}^{(k)} \\
\vdots & \vdots & \ddots & \vdots \\
-\frac{1}{2} \gamma_p\, t\, \rho_{0p}^{(k)} & -\frac{1}{2} \gamma_{p-1}\, t\, \rho_{0(p-1)}^{(k)} & \cdots &  0  \end{array} \right),
\label{Afirstorder}
\end{eqnarray}
where the dissipative evolution of site $k$ has been made explicit using the basis states $\left|0\right>_k$, $\left|1\right>_k, \left|2\right>_k, \ldots, \left|p\right>_k$, and $\rho_{mn}^{(k)}$ are the $p^{N-1} \times p^{N-1}$ matrices defined by $\rho_{mn}^{(k)} = {}_k{\left< n\right|} \rho \left|m\right>_k$. 
By analogously deriving higher order terms, it can be shown that the action of the dissipator $e^{\mathcal{L}_0 t} \rho$ is
\begin{equation}
e^{-i H_0 t} \left[ \bigotimes_{m \neq k} e^{\mathcal{L}_{0,d}^m t} \left( \begin{array}{ccccc}
\rho_{pp}^{(k)} & e^{-\frac{1}{2}(\gamma_{p-1}+\gamma_p) t} \rho_{p(p-1)}^{(k)}  & \cdots &  e^{-\frac{1}{2} \gamma_p t}\rho_{p0}^{(k)}  \\
e^{-\frac{1}{2}(\gamma_{p-1}+\gamma_p) t} \rho_{(p-1)p}^{(k)} & \rho_{(p-1)(p-1)}^{(k)} & \cdots &  e^{-\frac{1}{2} \gamma_{p-1} t}\rho_{(p-1)0}^{(k)} \\
\vdots & \vdots & \ddots & \vdots \\
e^{-\frac{1}{2} \gamma_p t} \rho_{0p}^{(k)} & e^{-\frac{1}{2} \gamma_{p-1} t} \rho_{0(p-1)}^{(k)}& \cdots &  \rho_{00}^{(k)}  \end{array} \right) \right] e^{i H_0 t}.
\label{Aactiondissippspec}
\end{equation}
The off-diagonal entries of the density matrix are seen to decay exponentially, a fact that is not altered by the action of the coherenct dynamics given by $H_0$, which is diagonal in  the product basis formed by single particle states $\left|0\right>_k$,$\left|1\right>_k, \left|2\right>_k,\ldots, \left|p\right>_k$. Therefore, the evolution under $\mathcal{L}_0$ becomes, at time scales much larger than the inverse of the dephasing rates, a projector $\mathcal{P}$ on the diagonal of $\rho$ in that same basis
\begin{equation}
\mathcal{P}\rho = \lim_{t\to\infty} e^{\mathcal{L}_0 t} \rho = \textrm{diag}(\rho),
\label{Aprojector}
\end{equation}
as happens in the case of just one Rydberg level \cite{lesanovsky2013}. The removal of all coherences leads to a diagonal density matrix, where each classically accessible configuration (e.g. $\left|0 0 1 0  0 2 0 3 \cdots 1\right>$) is given a certain probability of occurrence.

Using the projector operator $\mathcal{P}$ and its complement $\mathcal{Q} = 1-\mathcal{P}$, we can formulate the effective evoluton equation for the diagonal density matrix $\mu = \mathcal{P}\rho$ describing the slow evolution. To second order in $\mathcal{L}_1$, the general expression is given by
\begin{equation}
\partial_t \mu = \mathcal{P}\mathcal{L}_1\mu + \int_0^\infty dt \mathcal{P} \mathcal{L}_1 \mathcal{Q} e^{\mathcal{L}_0 t} \mathcal{Q} \mathcal{L}_1 \mu. 
\label{Aeffectivedynpre}
\end{equation}
In this case $\mathcal{P}\mathcal{L}_1\mu = 0$ and $\mathcal{Q} e^{\mathcal{L}_0 t} \mathcal{Q} \mathcal{L}_1 \mathcal{P} =   e^{\mathcal{L}_0 t} \mathcal{L}_1\mathcal{P}$. We next calculate the integrand in Eq. (\ref{Aeffectivedynpre}),
\begin{eqnarray}
&\mathcal{P} \mathcal{L}_1 e^{\mathcal{L}_0 t} \mathcal{L}_1 \mu &= - \mathcal{P} \left( \sum_{s s^\prime}  \sum_{km}  \Omega_{s^\prime} \Omega_s [\sigma_{s^\prime x}^{(k)}\,, e^{\mathcal{L}_0 t}  [\sigma_{s x}^{(m)}, \mu]]\right) \nonumber\\
& &= - \sum_{s} \sum_{k} \Omega_s^2  \, \mathcal{P} \left( \sigma_{sx}^{(k)} e^{\mathcal{L}_0 t} \left(\sigma_{sx}^{(k)} \mu\right) -\sigma_{sx}^{(k)} e^{\mathcal{L}_0 t} \left(\mu\, \sigma_{sx}^{(k)}\right)- e^{\mathcal{L}_0 t} \left(\sigma_{sx}^{(k)} \mu\right) \sigma_{sx}^{(k)} +  e^{\mathcal{L}_0 t}  \left(\mu\, \sigma_{sx}^{(k)}\right) \sigma_{sx}^{(k)}\right).
\label{Aintegrand}
\end{eqnarray}
We have used the fact that $e^{\mathcal{L}_0 t} $ does not shift matrix elements, and that the action of $\sigma_{sx}^{(k)} = \left|s\right>_k\!\left<0\right| + \left|0\right>_k\!\left<s\right|$ followed by that of $\sigma_{s^\prime x}^{(m)} = \left|s^\prime\right>_m\!\left<0\right| + \left|0\right>_m\!\left<s^\prime\right|$ can only produce non-zero diagonal elements if $s^\prime = s$ and $m = k$.

In the following, we explicitly work out the terms in  Eq. (\ref{Aintegrand}). We focus on the contribution corresponding to level $\left|1\right>$ for concreteness.
\begin{eqnarray}
&\sigma_{1x}^{(k)} e^{\mathcal{L}_0 t} \left(\sigma_{1x}^{(k)} \mu\right) &= \sigma_{1x}^{(k)} e^{\mathcal{L}_0 t}  \left( \begin{array}{cccc}
0  &  \cdots & 0 & 0 \\
\vdots  &  \ddots & \vdots & \vdots \\
0 & \cdots & 0 & \rho_{00}^{(k)} \\
0 & \cdots & \rho_{11}^{(k)} &  0  \end{array} \right) 
= \sigma_{1x}^{(k)} e^{-i H_0 t}  \left( \begin{array}{cccc}
0  &  \cdots & 0 &  0 \\
\vdots  &  \ddots & \vdots & \vdots \\
0  &  \cdots & 0 &  e^{-\frac{1}{2} \gamma_1 t}\rho_{00}^{(k)} \\
0  &  \cdots  & e^{-\frac{1}{2} \gamma_1 t} \rho_{11}^{(k)} &  0  \end{array} \right) e^{i H_0 t}\\ \nonumber
& &= \left( \begin{array}{ccc}
\ddots & \vdots & \vdots \\
\cdots & e^{-\frac{1}{2} \gamma_1 t} e^{i t \sum_m \left[V^1_{km} n_1^{(m)} + \sum_s V^{1 s}_{km} n_s^{(m)}\right]} \rho_{11}^{(k)} &  0 \\
\cdots & 0 &  e^{-\frac{1}{2} \gamma_1 t} e^{-i t \sum_m \left[ V^1_{km} n_1^{(m)} +  \sum_s V^{1 s}_{km} n_s^{(m)}\right]} \rho_{00}^{(k)}  \end{array} \right)
\end{eqnarray}

\begin{eqnarray}
&\sigma_{1x}^{(k)} e^{\mathcal{L}_0 t} \left(\mu\, \sigma_{1x}^{(k)}\right) &= \left( \begin{array}{ccc}
\ddots & \vdots & \vdots \\
\cdots & e^{-\frac{1}{2} \gamma_1 t} e^{i t \sum_m \left[ V^1_{km} n_1^{(m)} + \sum_s V^{1s}_{km} n_s^{(m)}\right]} \rho_{00}^{(k)} &  0 \\
\cdots & 0 &  e^{-\frac{1}{2} \gamma_1 t} e^{-i t \sum_m \left[V^1_{km} n_1^{(m)} + \sum_s V^{1s}_{km} n_s^{(m)}\right]} \rho_{11}^{(k)}  \end{array} \right)
\end{eqnarray}

\begin{eqnarray}
&e^{\mathcal{L}_0 t} \left(\sigma_{1x}^{(k)} \mu\right) \sigma_{1x}^{(k)} &= \left( \begin{array}{ccc}
\ddots & \vdots & \vdots \\
\cdots & e^{-\frac{1}{2} \gamma_1 t} e^{-i t \sum_m \left[ V^1_{km} n_1^{(m)} + \sum_s V^{1s}_{km} n_s^{(m)}\right]} \rho_{00}^{(k)}  &  0 \\
\cdots & 0 &  e^{-\frac{1}{2} \gamma_1 t} e^{i t \sum_m \left[ V^1_{km} n_1^{(m)} + \sum_s V^{1s}_{km} n_s^{(m)}\right]} \rho_{11}^{(k)}  \end{array} \right)
\end{eqnarray}

\begin{eqnarray}
&e^{\mathcal{L}_0 t}  \left(\mu\, \sigma_{1x}^{(k)}\right) \sigma_{1x}^{(k)} &= \left( \begin{array}{ccc}
\ddots & \vdots & \vdots \\
\cdots & e^{-\frac{1}{2} \gamma_1 t} e^{-i t \sum_m \left[ V^1_{km} n_1^{(m)} + \sum_s V^{1s}_{km} n_s^{(m)}\right]} \rho_{11}^{(k)} &  0 \\
\cdots & 0 &   e^{-\frac{1}{2} \gamma_1 t} e^{i t \sum_m \left[ V^1_{km} n_1^{(m)} + \sum_s V^{1s}_{km} n_s^{(m)}\right]} \rho_{00}^{(k)} \end{array} \right)
\end{eqnarray}

We will use $\mathcal{V}^k_s = \sum_m \left[ V^s_{km} n_s^{(m)} + \sum_{s^\prime\neq s} V^{ss^\prime}_{km} n_{s^\prime}^{(m)}\right]$ as shorthand to refer to the increment in the interaction energy that one has to pay for the excitation of atom $k$ to level $|s\rangle$. In the expressions above $\mathcal{V}^k_1$ appears in the oscillatory part of the diagonal elements of the matrix.

The term corresponding to $s=1$ in Eq. (\ref{Aintegrand}) is therefore
\begin{eqnarray} -\Omega_1^2  \left( \begin{array}{ccc}
\ddots & \vdots & \vdots \\
\cdots & 2\, e^{-\frac{1}{2} \gamma_1 t} \cos\left({\mathcal{V}^k_1 t}\right) \left[\rho_{11}^{(k)}-\rho_{00}^{(k)}\right]&  0 \\
\cdots & 0 &  2\, e^{-\frac{1}{2} \gamma_1 t} \cos\left(\mathcal{V}^k_1 t\right) \left[\rho_{00}^{(k)}-\rho_{11}^{(k)}\right] \end{array} \right),
\end{eqnarray}
and the contributions due to the other levels take an analogous form. Thus, Eq. (\ref{Aeffectivedynpre}) can be rewritten as
\begin{equation}
\partial_t \mu = -\sum_{s=1}^p \Omega_s^2 \int_0^\infty dt\, \sum_k 2\, e^{-\frac{1}{2} \gamma_s t} \cos\left({\mathcal{V}^k_s t}\right) \left[ \mathcal{I}_s^{(k)} \mu - \sigma_{sx}^{(k)} \mu\sigma_{sx}^{(k)}\right] = \sum_{s=1}^p  \sum_k \frac{ 4 \Omega_s^2/\gamma_s}{1 + (2 \mathcal{V}_s^k/\gamma_s)^2}  \left[\sigma_{sx}^{(k)} \mu\sigma_{sx}^{(k)} -  \mathcal{I}_s^{(k)} \mu\right],
\label{Aeffectivedynpreintpspec}
\end{equation}
where the projection operator $\mathcal{I}_s^{(k)} = n_{s}^{(k)} + |0\rangle \langle 0|$ cancels all the elements in $\mu$ that do not correspond to the ground state or $|s\rangle$ at site $k$. 
The effective dynamics is therefore given by
\begin{equation}
\partial_t \mu = \sum_{s=1}^p \frac{4\Omega_s^2}{\gamma_s} \sum_k \Gamma_{s}^{(k)} \left[\sigma_{sx}^{(k)} \mu \sigma_{sx}^{(k)} -  \mathcal{I}_s^{(k)} \mu \right],
\label{Aeffectiveintpspec}
\end{equation}
with rates for a transition $|0\rangle \to |s\rangle$ or $|s\rangle \to |0\rangle$ 
\begin{equation}
\Gamma_{s}^{(k)} = \frac{1}{1 + \left[\frac{2}{\gamma_{s}} \sum_m \left(V^{s}_{km} n_{s}^{(m)} + \sum_{s^\prime \neq s} V^{ss^\prime}_{km} n_{s^\prime}^{(m)}\right)\right]^2}.
\label{Apreratesintpspec}
\end{equation}
To make explicit the power-law interactions, it is useful to refer to the atomic spatial arrangement in terms of reduced position vectors $\hat{\bf r}_k = {\bf r}_k/a$, where $a$ is the lattice constant. We define an intra-level interaction parameter $R_{s}=a^{-1}[2C^{s}_\alpha/\gamma_{s}]^{1/\alpha}$ (for interactions between atoms in the same level, $V^{s}_{km} = C_\alpha^{s} n^{(k)}_{s}   n^{(m)}_{s}/a^\alpha|\hat{\bf r}_k - \hat{\bf r}_m|^\alpha$), and an inter-level interaction parameter $R_{s}^{s s^\prime}=a^{-1}[2C^{s s^{\prime}}_\alpha/\gamma_{s}]^{1/\alpha}$ (for interactions between atoms in different levels, $V^{s s^\prime}_{km} = C_\alpha^{s s^{\prime}} n^{(k)}_{s}   n^{(m)}_{s^\prime}/a^\alpha|\hat{\bf r}_k - \hat{\bf r}_m|^\alpha$).  The (inverse) rates can then be written as 
\begin{equation}
\frac{1}{\Gamma_{s}^{(k)}} = 1 + \left[\sum_m \frac{(R_{s})^\alpha\,n_{s}^{(m)}+ \sum_{s^\prime \neq s}(R^{ss^{\prime}}_{s})^\alpha n_{s^\prime}^{(m)}}{|\hat{\bf r}_k - \hat{\bf r}_m|^\alpha}\right]^2.
\label{Aratesintpspec}
\end{equation}
 In some cases, the interaction exponent $\alpha$ could be different depending on the atomic levels involved. This more general case can be easily worked out from  Eq. (\ref{Apreratesintpspec}), but here we will assume that $\alpha$ is the same for all level pairs.

\section{Appendix B. Phenomenology of a four-component dissipative Rydberg gas}

While the derivation of the effective equations of motion is valid for any number of species, in the numerical results reported in the main text we focus on the two-component case, $p=2$. However, both the main observations on the phenomenology and the theoretical arguments given there can be extended without great difficulty to the $p>2$ case. In this section we briefly report some results for $p=4$. For the sake of simplicity, we again use a somewhat idealized parameter choice according to which all the intra-level interaction parameters, which we collectively denote as $R$, are equal to one another, while all the inter-level interaction parameters, denoted as $R_c$, are also equal among themselves. We focus on the $R_c > R$ case, where homogeneous domains emerge, as it gives the richest phenomenology. More specifically, we consider $R=2$ and $R_c=8$, which coincides with the parameter choice used in the main text.

\begin{figure}[h]\hspace{0.5cm}
\includegraphics[scale=0.19]{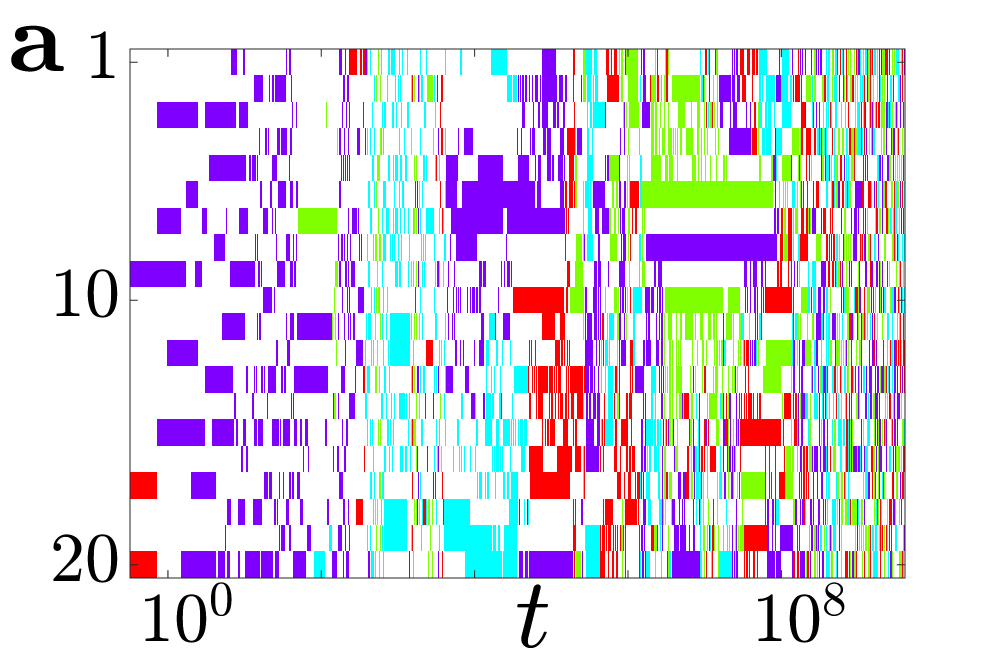}\hspace{0.2cm}
\includegraphics[scale=0.19]{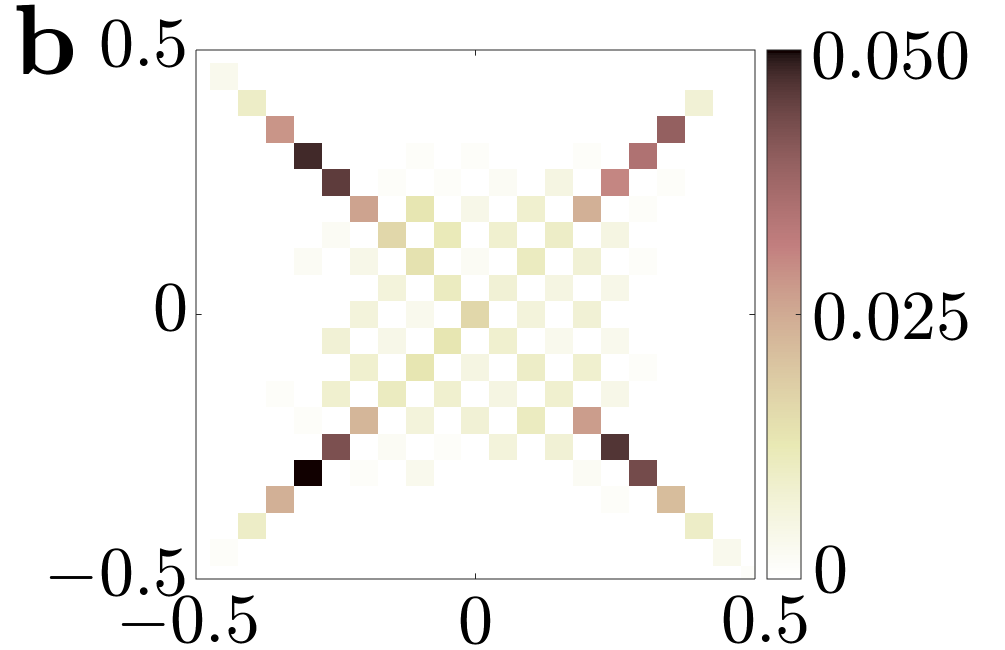}
\caption{{\sf \bf Individual trajectory and probability distribution of the order parameter $P_{4+}$ at $t=10^2$ for a $p=4$ system with $R=2$ and $R_c=8$.}
(a) Representative trajectory. The color coding is such that red corresponds to state $|1\rangle$, green to $|2\rangle$, cyan to $|3\rangle$, magenta to $|4\rangle$ and white to ground state atoms. (b) Probability distribution of $P_{4+}(t)$ at $t=10^2$. The histogram is based on $1000$ kinetic Monte Carlo realizations. }\label{fig5}
\end{figure}

In Fig. \ref{fig5} (a), we see one representative trajectory of a system of $N=20$ atoms with van der Waals interactions. The color coding is such that red corresponds to state $|1\rangle$, green to $|2\rangle$, cyan to $|3\rangle$, magenta to $|4\rangle$ and white to ground state atoms. While there are some initial excitations to $|1\rangle$, they finally de-excite and are replaced by excitations to $|4\rangle$, which by then has become the dominant component. The large homogeneous $|4\rangle$-domain that emerges is later replaced by a $|3\rangle$-domain. At longer times, two domains, corresponding to $|1\rangle$ and $|4\rangle$ coexist. Eventually, when stationarity is approached, the system undergoes a strong mixing of all the components. 

To quantify the emerging dynamical order we focus on a complex order parameter that is an extension of the real order parameter $P_+$ that was proposed in the main text for $p=2$ [see Eq. (\ref{pplus})]. It is defined as follows
\begin{equation}
P_{4+}(t) = \frac{1}{N}\sum_{k=1}^N \sum_{s=1}^4 \exp{ i \pi \left( \frac{1+2(s-1)}{4}\right)}\, n_s^{(k)}(t).
\label{P+4}
\end{equation}
This order parameter, which has been inspired by the theory of the Potts model \cite{Wu1982}, can be easily extended to any number of components $p$. In Fig. \ref{fig5} (b) we show $P_{4+}(t)$ at $t=10^2$, which corresponds to the time at which most of the trajectories inspected still show one domain that spans the whole chain. The existence of as many maxima as there are excited levels, all of them quite distant from the origin, indeed indicates that the formation of large domains of the kind seen in Fig. \ref{fig2} (c) and (d) of the main text for $p=2$ occurs generically in systems with a larger number of components as well. As in the two-component case [main text, Fig. \ref{fig4} (d)], the four peaks reach the saturation value of $|P_{4+} |\simeq 0.5$ at later times, and eventually subside into a unimodal distribution centered around the origin when the system approaches the stationary state.

\section{Acknowledgements}

We  thank  Beatriz Olmos and Weibin Li for insightful discussions. The research leading to these results has received funding from the European Research Council under the European Union's Seventh Framework Programme (FP/2007-2013) / ERC Grant Agreement No. 335266 (ESCQUMA), the EU-FET grant HAIRS 612862 and from the University of Nottingham. Further funding was received through the H2020-FETPROACT-2014 grant No.  640378 (RYSQ). We also acknowledge financial support from EPSRC Grant no.\ EP/M014266/1. Our work has benefited from the computational resources and assistance  provided  by the University of Nottingham High Performance Computing service.

%


\end{document}